\documentclass[twocolumn,tighten,trackchanges,apj]{aastex63}

\usepackage{apjfonts}
\usepackage{amsmath}
\usepackage{amsfonts}
\usepackage{amssymb}
\usepackage{mathtools}

\DeclareMathOperator\erfc{erfc}
\DeclareMathOperator\heavi{H}

\renewcommand{\vec}{\textbf}

\newcommand{\nostar}{\vphantom{*}}
\newcommand{\notwo}{\vphantom{2}}

\newcommand{\Sec}[1]{Section~\ref{#1}}
\newcommand{\lperp}{l_{\perp}}
\newcommand{\lpara}{l_{\parallel}}
\newcommand{\BE}{\begin{equation}}
\newcommand{\EE}{\end{equation}}
\newcommand{\zero}{_{\mathrm 0}}
\newcommand{\Eq}[1]{Equation~(\ref{#1})}
\newcommand{\Eqs}[2]{Equations~(\ref{#1}) and~(\ref{#2})}

\newcommand{\RL}{R_{\mathrm L}}
\newcommand{\HRL}{\hat{R}_{\mathrm L}}
\newcommand{\RLL}{\RL/\lperp}
\newcommand{\Fig}[1]{Figure~\ref{#1}}
\newcommand{\Figs}[2]{Figures~\ref{#1} and~\ref{#2}}

\newcommand{\Bzero}{B_{0}}
\newcommand{\dbob}{b/\Bzero}

\shortauthors{Snodin et al.}

\begin{document}

\title{Energetic Particle Perpendicular Diffusion: Simulations and Theory in Noisy Reduced Magnetohydrodynamic Turbulence}

\author[0000-0001-7551-3511]{A.~P.~Snodin}
\affiliation{Department of Mathematics, Faculty of Applied Science,
King Mongkut's University of Technology North Bangkok, Bangkok 10800, Thailand}
\affiliation{Department of Materials and Production Technology Engineering,
Faculty of Engineering, King Mongkut's University of Technology North Bangkok,
Bangkok 10800, Thailand}
\affiliation{Department of Physics, Faculty of Science, Mahidol University,
Bangkok 10400, Thailand}

\author{T.~Jitsuk}
\altaffiliation{Present address: Department of Physics, University of Wisconsin-Madison, Madison, WI 53706, USA}
\affiliation{Department of Physics, Faculty of Science, Mahidol University,
Bangkok 10400, Thailand}

\author[0000-0003-3414-9666]{D.~Ruffolo}
\affiliation{Department of Physics, Faculty of Science, Mahidol University,
 Bangkok 10400, Thailand}

\author[0000-0001-7224-6024]{W.~H.~Matthaeus}
\affiliation{Bartol Research Institute and Department of Physics and Astronomy,
University of Delaware, Newark, DE 19716, USA}
\correspondingauthor{D.~Ruffolo}
\email{david.ruf@mahidol.ac.th}


\begin{abstract}
The transport of energetic charged particles (e.g., cosmic rays) in turbulent magnetic fields is usually characterized in terms of the diffusion parallel and perpendicular to a large-scale (or mean) magnetic field.
The nonlinear guiding center theory (NLGC) has been a prominent perpendicular
diffusion theory. A recent version of this theory, based on random ballistic spreading of magnetic field lines and a backtracking correction (RBD/BC), has shown good agreement with test particle simulations for a two-component magnetic turbulence model. The aim of the present study is to test the generality of the improved theory by applying it to the noisy reduced magnetohydrodynamic (NRMHD) turbulence model, 
determining perpendicular diffusion coefficients that are compared with those from the field line random walk (FLRW) and unified nonlinear (UNLT) theories and our test particle simulations.
The synthetic NRMHD turbulence model creates special conditions for
energetic particle transport, with no magnetic fluctuations at higher parallel
wavenumbers so there is no resonant parallel scattering if the particle Larmor radius $\RL$ is even slightly smaller than the minimum resonant scale.
This leads to non-monotonic variation in the parallel mean free path $\lambda_\parallel$ with $\RL$.
Among the theories considered, only RBD/BC matches simulations within a factor of two over the range of parameters considered.
This accuracy is obtained even though the theory depends on $\lambda_\parallel$ and has no explicit dependence on $\RL$. 
In addition, the UNLT theory often provides accurate results and even the FLRW limit provides a very simple and reasonable approximation in many cases.

\end{abstract}





\section{INTRODUCTION}

\begin{figure}
\plotone{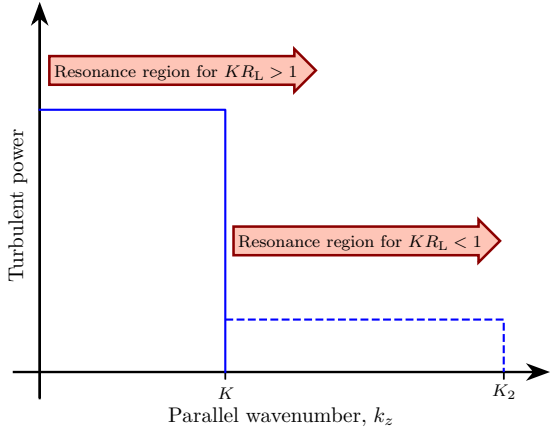}
\caption{Parallel power spectrum of the noisy reduced magnetohydrodynamic (NRMHD) turbulence models used here, for positive parallel wavenumber $k_z$.  According to quasilinear scattering theory \citep{Jokipii66}, 
particles can resonantly scatter in pitch angle due to fluctuations at $k_z\geq 1/\RL$, where $\RL$ is the Larmor radius, as indicated
by labeled arrows.  Solid line: Original model spectrum of \cite{RuffoloMatthaeus13} with sharp cutoff at $k_z=K$.  For this model, if $K\RL>1$, some particles undergo resonant scattering and others do not, i.e., there is a resonance gap.  For $K\RL<1$, there is a complete lack of resonant scattering.  In either case, nonlinear interactions allow some particle scattering.  Dashed line: We also consider adding fluctuation power at a lower level up to $k_z=K_2$.  In this case, 
there is some resonant scattering for a wider range $K_2\RL>1$.  In the present work, simulations using these NRMHD models allow us to explore the effects of the resonance gap or lack of resonant scattering on parallel and perpendicular diffusion, in comparison with theories of perpendicular diffusion.
\label{fig:orient}} 
\end{figure}
Modeling energetic charged particle (or cosmic ray) transport, and interpreting various astrophysical and laboratory plasma phenomena requires a detailed description of charged particle behavior in turbulent magnetic fields. This behavior is usually expected to be spatially {\it diffusive} on length scales larger than the coherence length of the turbulent magnetic field \citep{MeyerEA56,Parker65}. In the presence of an identifiable large-scale (or mean) field, one is usually interested in how particles diffuse in the directions parallel or perpendicular to this field direction.  
The present work focuses on perpendicular diffusion, which is directly relevant to understanding the distribution of cosmic rays in the heliosphere \citep[e.g.,][]{StraussEA12}, relating to a key health hazard for extended human space missions \citep{Knipp11}, as well as solar modulation of the cosmic ray flux according to the $\sim$11-year solar activity cycle \citep{Forbush54} and $\sim$22-year solar magnetic cycle \citep{JokipiiThomas81}.
Perpendicular and parallel diffusion are also key to diffusive shock acceleration of energetic particles \citep{Krymskii77,AxfordEA77,Bell78,Drury83}
and are relevant to the distribution of cosmic rays in our Galaxy \citep{RecchiaEA16} and other galaxies \citep{Heesen21}.

Since particles tend to gyrate along magnetic field lines, their perpendicular transport is expected to depend on the random walk of those field lines. This is the basis of the field line random walk (FLRW) theory \citep{Jokipii66}, where the perpendicular diffusion coefficient $\kappa_\perp$ is directly related to the magnetic field line diffusion coefficient $D$. However, numerical test particle simulations, an important tool in verifying transport theories, have shown limited agreement with the FLRW theory. In addition to the random walk of magnetic field lines, the perpendicular transport is also thought to depend on both parallel scattering (that causes the particle to change direction relative to the large scale magnetic field) and the nature of the transverse magnetic field structure that a particle experiences. The nonlinear guiding center (NLGC) theory \citep{MatthaeusEA03} was developed to allow for the decorrelation of the particle trajectories after the parallel scattering to contribute to the perpendicular diffusion coefficient. The theory has shown good agreement with both observations and computer simulations \citep[e.g.,][]{BieberEA04,RuffoloEA2008} and influenced several related theories \citep[see][and references therein]{Shalchi09book,Shalchi2010}.

In the original NLGC theory it was assumed that particle guiding centers have a diffusive behavior, which results in a nonlinear integral for the perpendicular diffusion coefficient. A recent alternative theory \citep{RuffoloEA12}, that we refer to as NLGC (RBD/BC), assumes that the field lines followed by particles spread ballistically at early times so that guiding centers undergo random ballistic decorrelation (RBD). This theory also includes a backtracking correction (BC) that accounts for the reduction in perpendicular diffusion due to parallel scattering, which causes a particle to reverse its direction along the field line.
The RBD/BC theory was originally tested using direct numerical simulations for the two component magnetic turbulence model that combines a two dimensional component that varies perpendicular to the mean field and a slab component that varies in the parallel direction, which is widely used to model magnetic fluctuations in the solar wind \citep{MatthaeusEA90}. The RBD/BC theory and test particle simulations were found to have improved agreement over the original NLGC theory. 

In this study, we aim to test the generality of the NLGC (RBD/BC) theory by applying it to the noisy reduced magnetohydrodynamic turbulence (NRMHD) model \citep[][]{RuffoloMatthaeus13}, which will be described in the next section.
NRMHD, a synthetic model of homogeneous turbulence, is interesting for particle transport in that its power spectrum has a sharp cutoff at $k_z=K$ in the parallel wavenumber $k_z$, as shown in \Fig{fig:orient} (solid trace). According to the quasilinear scattering theory \citep{Jokipii66},  
particles at pitch angle cosine $\mu$ resonate with fluctuations at $k_z=1/(|\mu|\RL)$, where $\RL$ is the Larmor radius.  Since $1/|\mu|\geq1$ and NRMHD only has fluctuations at $k_z\leq K$,
there is no parallel resonant scattering whenever $K \RL < 1$.  
In turn, this implies that an ensemble of particles at sufficiently small $\RL$ has an infinite parallel mean free path according to quasilinear theory.
In actuality, parallel diffusion can be enabled by nonlinear effects that allow changes in pitch angle and reversal of propagation direction, despite the resonance gap (for $\RL>1/K$) or complete lack of resonance (for $\RL<1/K$).
Thus in the NRMHD field, we expect that the parallel mean free path $\lambda_\parallel$ can be very long, but not infinite, and its actual value is sensitively determined by nonlinear effects.\footnote{While here we consider a resonance gap of varying width or complete lack of resonance, it is worth noting that resonant scattering can be broadened by dynamical effects \citep{BieberEA94}, perpendicular diffusion \citep{ShalchiEA04-WNLT}, or second-order effects \citep{Shalchi05-SOQLT}.}

This is of particular relevance to NLGC-based theories of perpendicular diffusion, such as RBD/BC and the unified nonlinear theory \citep[UNLT;][]{Shalchi2010}, because these theories determine the perpendicular diffusion coefficient $\kappa_\perp$ or perpendicular mean free path $\lambda_\perp=3\kappa_\perp/v$ (where $v$ is the particle speed) from an input value of $\lambda_\parallel$.
\citet{ShalchiHussein2014}
found that the original NLGC theory does not provide the correct dependence of $\lambda_\perp$ on $\lambda_\parallel$ for extremely long $\lambda_\parallel$ values that can result in NRMHD turbulence, in comparison with their test particle simulations with the ratio of parallel to perpendicular correlation lengths of the turbulence set to  $\lpara/\lperp=\pi/2$ and the ratio of the rms magnetic fluctuation amplitude to the mean field set to $b/B_0=1$, while the UNLT theory provides a good match to simulation results for those parameter values.

The RBD/BC theory has corrected this problem with the original NLGC theory, yielding $\lambda_\perp$ values that saturate to a constant level as $\lambda_\parallel\to\infty$.
The present work aims to test whether the RBD/BC theory can accurately determine $\lambda_\perp$ even when the input $\lambda_\parallel$ is sensitively determined by nonlinear parallel transport effects and may take on extreme values, for various 
combinations of $\lpara/\lperp$ and $b/B_0$.
Here we perform computer simulations of particle transport in the NRMHD model of magnetic turbulence to determine the  parallel and perpendicular diffusion coefficients for various model parameters, which are then used to validate RBD/BC and UNLT as well as a composite model \citep{Shalchi2019} and the classic field line random walk (FLRW) theory \citep{Jokipii66}.


\section{MAGNETIC FIELD MODEL}\label{model}


\subsection{The NRMHD Model}

The NRMHD model \citep{RuffoloMatthaeus13} is a synthetic model of homogeneous turbulence designed to emulate the power spectrum found in dynamical simulations based on the reduced magnetohydrodynamic (RMHD) equations for nearly transverse fluctuations with a slow variation parallel to the mean magnetic field, which were developed to describe tokamak plasmas \citep{Strauss76} and later widely employed to model fluctuations in solar coronal magnetic loops \citep[e.g.,][]{LongcopeSudan94}. 
In the NRMHD model, the total magnetic field is expressed as
\BE
\vec{B} = B\zero\hat{\vec{z}} + \vec{b}(\vec{x}),
\EE
where $B\zero\hat{\vec{z}}$ is the background mean field.
This model considers transverse turbulence with 
$\vec{b}\perp\hat{\vec{z}}$, based on the physical expectation that fluctuations are enhanced in the transverse directions (i.e., with variance anisotropy) when there is a strong mean field with $B_0\gtrsim b$, where $b^2\equiv\langle\vec{b}^2\rangle$. 
The fluctuating field $\vec{b}$ can be expressed in terms of a scalar potential function $a$ as $\vec{b}(\vec{x}) = \nabla \times \left[a(\vec{x})\hat{\vec{z}}\right]$, which in terms of wavevectors \vec{k}, can be written as
\BE \label{eq:bk}
\vec{b}(\vec{k}) = -i\vec{k} \times \left[a(\vec{k})\hat{\vec{z}}\right].
\EE
For NRMHD we take
\BE \label{eq:a}
a(\vec{k}) = \frac{a^{\rm 2D}(k_x,k_y)}{\sqrt{4\pi K}}e^{i\phi(\vec{k})}\heavi(K - |k_z|),
\EE
where $\phi(\vec{k})\in[0,2\pi)$ is a random phase that is independent for each $\vec{k}$, $\heavi$ is the unit step function and $K$ is the maximum wavenumber in the $z$-direction.
For convenience we adopt turbulence that is axisymmetric about the mean magnetic field. 
Viewed as an extension of a 2D spectrum to three dimensions of wavenumber space, it is similar to a spectral component used by \cite{WeinhorstShalchi2010}, but NRMHD has the interesting property of a sharp cutoff in $k_z$ that leads to a resonance gap or a complete lack of resonance for parallel scattering as described in Section 1.

Then we can write the three-dimensional power spectra $S_{xx}(\vec{k}) = \langle b_x^{\nostar}(\vec{k})b_x^{*}(\vec{k}) \rangle$ and $S_{yy}(\vec{k}) = \langle b_y^{\nostar}(\vec{k})b_y^{*}(\vec{k}) \rangle$, where $^{*}$ denotes the complex conjugate, as
\BE
S_{xx}(\vec{k}) = \frac{k_y^2A(k_\perp)}{4\pi K}\heavi(K - |k_z|), \quad S_{yy}(\vec{k}) = \frac{k_y^2A(k_\perp)}{4\pi K}\heavi(K - |k_z|),
\EE
where $A(k_\perp)$ is the power spectrum of $a^{\rm 2D}$, with $k_\perp = \sqrt{k_x^2 + k_y^2}$.
Note that the total magnetic power spectrum $S(k_\perp)=S_{xx}(\vec{k}) + S_{yy}(\vec{k})$ is axisymmetric and a function only of $k_\perp$. 
The choice of the normalization factor in \Eq{eq:a} allows one to write the total fluctuation energy as
\BE \label{eq:b2}
b^2 = \int S(k_\perp) d\vec{k} = \int_0^{\infty} k_{\perp}^3 A(k_\perp) d k_\perp,
\EE
which is independent of the maximum parallel wavenumber $K$. 

We make a specific choice for $A(k_\perp)$ in order to implement the model in computer simulations and evaluate theoretical expressions. We take
\BE \label{eq:A}
A(k_\perp) = \frac{A\zero}{\left[1 + (\lambda k_\perp)^2 \right]^{7/3}},
\EE
where $A\zero = 8 b^2 \lambda^4/9$ is required to satisfy \Eq{eq:b2} and $\lambda$ is a perpendicular bend-over scale.
Viewing $S(k_\perp)$ as a power spectrum of axisymmetric 2D turbulence, this form leads to $S\propto k_\perp^2$ at small $k_\perp$, to model fluctuations in the energy-containing range, a dependence that is consistent with asymptotic homogeneity \citep{MatthaeusEA2007}.
It also provides an omnidirectional 2D energy spectrum 
\begin{eqnarray}
    {\cal E}(k_\perp)&=&2\pi k_\perp S(k_\perp)\nonumber\\
    &\propto& k_\perp^3A(k_\perp)\nonumber\\
    &\propto& k_\perp^{-5/3} \qquad\qquad (k_\perp\gg1/\lambda)
\end{eqnarray}
at high $k_\perp$, to model an inertial range of turbulence that is consistent with Kolmogorov theory \citep{Kolmogorov41a,Batchelor70}.

The model has a correlation length $\lpara = \pi / (2K)$ along $z$ and a total perpendicular correlation length \citep{MatthaeusEA2007}
\BE \label{eq:lperp}
\lperp = \frac{\int k_\perp A(k_\perp) d\vec{k}_\perp}{\int k_\perp^2 A(k_\perp) d\vec{k}_\perp} =  \frac{\int_0^{\infty} k_\perp^2 A(k_\perp) d k_\perp}{\int_0^{\infty} k_\perp^3 A(k_\perp) d k_\perp}.
\EE
From this expression one finds that $\lambda \approx 2.678 \lperp$. The model can be characterised by the magnetic Kubo number,
\BE
R=\frac{b}{B\zero}\frac{\lpara}{\lperp},
\EE
where $b=\sqrt{\langle b^2 \rangle}$ is the rms fluctuation strength. 

The conditions for arriving at RMHD as a suitable dynamical model in a plasma have been investigated in some detail \citep{Montgomery82,OughtonEA17}. 
In general there is some subtlety in simultaneously obtaining the conditions of incompressibility, variance anisotropy and spectral anisotropy, thus implying that RMHD is likely more applicable to, say, solar coronal loops than it is to the interplanetary medium. Here we do not investigate the applicability of the model for particular plasmas, but simply adopt RMHD as given. 
We consider here the range $0.3 \le R\le10$, which is roughly compatible with constraints of RMHD theory \citep[as detailed in, e.g.,][]{OughtonEA17}.
Indeed, the NRMHD model is called ``noisy'' in the sense that it can be tuned to transition between quasi-2D turbulence, with $\lpara\gg\lperp$, as envisioned for RMHD, to a more slab-like configuration, with $\lpara\lesssim\lperp$, effectively adding noise to the RMHD model \citep{RuffoloMatthaeus13}.


\subsection{Small Scale Extension}
An interesting aspect of NRMHD in the context of test particle simulations is that for moderate values of the particle Larmor radius $\RL$ there is a large gap in the range of pitch angles for which the particle can undergo resonant scattering, which is a consequence of the form of the magnetic power spectrum and its sharp cutoff at $|k_z| = K$. In dynamical RMHD simulations, such as those presented in \cite{SnodinEA13b}, or more generally in reality, one expects some {\it finite} power at larger $k_z$ that will lead to some resonant scattering for particles of low $\RL$ (or energy). To extend the NRMHD model
to include some finite power at larger $k_z$, we add a small fraction of the total magnetic energy and spread it uniformly to $K_2$, a much larger maximum $k_z$ value, as shown by the dashed line in \Fig{fig:orient}. More precisely, we replace $\heavi(K - |k_z|)$ in \Eq{eq:a} with
\BE \label{eq:extension}
\Theta(k_z) =
\begin{cases}
1, & |k_z| \leq K \\
a_1, & K < |k_z| \leq K_2 \\
0, & |k_z| > K_2,
\end{cases}
\EE
where we take $a_1=0.01$, which is sufficient to obtain a noticeable change in parallel transport for the test particle simulations used here. Unless $K_2$ is many orders of magnitude larger than $K$, the magnetic energy at wavenumbers beyond $|k_z|=K$ will be negligible, and so $\lpara$ will be essentially unchanged. The effect on the theories presented later would also be negligible, and so we do not account for this extension when evaluating any theoretical expressions.


\subsection{Numerical Implementation}

In our computer simulations, magnetic field realizations are produced on a three-dimensional grid of
size $N_x \times N_y \times N_z$. We construct the field on a grid in wavevector space by first drawing a phase $\phi$ from a random uniform distribution for each distinct wavevector for which \Eq{eq:a} is non-zero, i.e., where $|k_z| \leq K$, enforcing ${a}(\vec{k})={a}^{*}(-\vec{k})$ so that the magnetic field will be real-valued. 
We take $a^{\mathrm 2D}=\sqrt{A(k_\perp)}$.
We then construct $\vec{b}(\vec{k})$ according to Equations (\ref{eq:bk}) and (\ref{eq:a}), and apply an inverse FFT to each component to obtain $\vec{b}(\vec{x})$ on a periodic grid. The strength of the random field is controlled by re-normalizing the field to have the desired rms fluctuation strength $b$. The length scales $\lpara$ and $\lperp$ are not assumed to be as given above but are evaluated through discrete sums over the wavevector space. The finite box size means that the limits of integration in \Eq{eq:lperp} are effectively restricted to a range of perpendicular wavenumbers, which means that $\lambda$ needs to be adjusted for given perpendicular dimensions to obtain the desired $\lperp$. Multiple realizations are constructed for each computer simulation in order to average over fields with the same statistical properties. Linear interpolation is used to obtain the magnetic field at a point between the grid points.

Rather than constructing the magnetic field on a discrete mesh, one could alternatively construct a continuous field with a finite sum of Fourier components in configuration space, as was done in \citet{GiacaloneJokipii99} and \cite{ShalchiHussein2014}.
With such a construction, the magnetic field is evaluated only at points near to a particle trajectory and the smallest scale fluctuations are essentially perfectly resolved, with no interpolation effects. However, a potential limitation is that the density of modes near resonant wavenumbers may be too low to achieve realistic resonant scattering \citep{MaceEA12}.
For some of the cases presented here, particularly when finite grid resolution effects may have been important, we repeated our simulations with such a field, taking a uniform distribution of wavenumbers in the parallel direction and logarithmically spaced $k_{\perp}$. The results were essentially identical to those obtained with the magnetic field constructed on a grid, while being much more computationally expensive to obtain.


\section{COMPUTER SIMULATIONS}\label{sec:simulations}


\subsection{Test Particle Simulations} \label{sec:testparticles}
We perform test particle simulations in realizations of NRMHD turbulence generated as described in Section~\Ref{model}.
We take a three-dimensional periodic box with perpendicular dimensions $L_x=L_y=30$, a perpendicular bend-over scale $\lambda=2.495425$, and $N_x=N_y=512$ grid points, which yields $\lperp \approx 1$ and a perpendicular grid resolution of $\delta x = \delta y = 0.05859$. 
In other words, we express all lengths in units of $l_\perp$.
Note that the value of $\lambda$ above has been obtained by an iterative process that aims to achieve $l_{\perp}=1$ by replacing the continuous integrals in \Eq{eq:lperp} with discrete sums over the wavevectors that exist in the computational grid. 

We also explore the effect of the aspect ratio $l_\parallel/l_\perp$ on the simulation results and the accuracy of the theories.
We consider a range from $\lpara/l_\perp=10$, a typical value used for RMHD simulations, to $\lpara/l_\perp=1$ as considered by \citet{ShalchiHussein2014}.
In the parallel direction, we take $N_z=4096$ grid points and use the box length $L_z$ to vary the parallel length scale $\lpara$. 
For the wavenumber range with non-zero turbulence power, $|k_z|\leq K$,
we use $64$ discrete modes each for the positive and negative $k_z$ values, which ensures that the smallest scale fluctuation at $k_z=K$ is well resolved. We also include a mode at $k_z=0$. For $L_z=258$ this configuration of modes for $\lpara\approx 1$ gives $K=1.571$ and a parallel grid resolution of $\delta z =0.06299$. We also consider the cases $l_{\parallel}\approx 3$ and $l_{\parallel}\approx 10$, where we take $L_z=774$ $(K=0.5236, \delta z=0.1890)$ and $L_z=2580$ $(K=0.1571,\delta z=0.6299)$, respectively.

For each particle in a realization we solve the (dimensionless) Newton-Lorentz equations for the particle trajectory $\vec{x}(t)$ and velocity $\vec{v}(t)$,
\BE
\frac{d\vec{v}}{dt} = \alpha \vec{v} \times \vec{B}\left[\vec{x}(t)\right], \quad \frac{d\vec{x}}{dt} = \vec{v},
\EE
where $\alpha=qB\zero\lperp/(\gamma m v\zero)$, with $q$ the particle charge, $\gamma$ the Lorentz factor, $m$ the particle rest mass and $v\zero$ the unit of velocity. Note that these equations imply that $|\vec{v}|$ is constant, or in other words, the particle energy is conserved. For simplicity we take $|\vec{v}|=1$ in units of $v\zero$, $B\zero=1$ and then
vary the parameter $\alpha$ to control the dimensionless Larmor radius $\HRL \equiv \RLL = \gamma m v\zero / (|q|B\zero)$.
The other important parameter, the relative fluctuation strength $b/B\zero$, is controlled by normalizing the magnetic field realization to obtain the desired rms fluctuation $b$. We typically use $200$ realizations for each set of parameters, with $100$ particles per realization. The initial particle locations are distributed uniformly over the box and their velocity vectors are distributed isotropically. The particle trajectories are integrated 
numerically using the 8th-order adaptive Runge-Kutta method of \cite{HairerEA93} (DOP853, using Dormand \& Prince coefficients), up to a time of $800-50,000 t_{\mathrm L}$, where $t_{\mathrm L}=2\pi \RL/v$, depending on the time taken for the particles to become diffusive.

The asymptotic diffusion coefficient in the $x$-direction is usually defined as
\BE \label{eq:kappaxx}
\kappa_{xx} \equiv \lim_{t\to \infty} \frac{\langle (\Delta x)^2\rangle}{2t},
\EE
and similarly for the $y$ and $z$-directions. Here $\Delta x = x(t) - x(0)$ is the displacement of the particle in the $x$-direction from its starting point after time $t$ and the angle brackets denote an ensemble average over particles and magnetic field realizations. In practice we need to take the diffusion coefficient at a sufficiently large time and use
\BE \label{eq:running}
\kappa_{xx}(t) = \frac{1}{2}\frac{d\langle (\Delta x)^2\rangle}{dt}(t),
\EE
which is identical to \Eq{eq:kappaxx} in the limit of large $t$, but converges more rapidly to a constant diffusion coefficient. Because we are using axisymmetric turbulence, we expect that $\kappa_{xx} = \kappa_{yy}$, so use the difference between
these quantities as a measure of the error. In addition, we combine the statistics to obtain a perpendicular diffusion coefficient $\kappa_{\perp} = (\kappa_{xx} + \kappa_{yy})/2$.


\subsection{Magnetic Field Line Tracing} \label{sec:FLT}
For each set of realizations we also trace magnetic field lines by integrating
\BE \label{eq:FL}
\frac{d \vec{x}^{\mathrm FL}}{dz} = \frac{\vec{b}(\vec{x}^{\mathrm FL})}{B\zero}
\EE
for a sufficiently large number of field lines that start at locations distributed uniformly over the box. The collection of field lines allows us to obtain
\BE \label{eq:D}
D_{xx} = \lim_{z \to \infty} \frac{1}{2}\frac{d\langle (\Delta x^{\mathrm FL})^2\rangle}{dz},
\EE
which is the magnetic field line diffusion coefficient in the $x$-direction. Since our fluctuations are axisymmetric, we
combine this with the diffusion coefficient in the $y$-direction to obtain the perpendicular diffusion coefficient, $D = (D_{xx} + D_{yy})/2$. Note that the diffusion of magnetic field lines is controlled solely by the magnetic Kubo number $R$ and so we could have simply taken the results from \cite{SnodinEA13b}, which are very similar. However, here we use the same realizations as in the test particle simulations for consistency. The theoretical expressions of \cite{RuffoloMatthaeus13} for $D$ also yield very similar results, particularly those suited to the low $R$ regime that we use in the present work.


\section{THEORIES FOR PERPENDICULAR DIFFUSION}

Here we first present the theories that we apply to the NRMHD turbulence model.
The asymptotic particle diffusion coefficient in \Eq{eq:kappaxx} can equivalently be expressed in the Taylor-Green-Kubo (TGK) form
\BE \label{eq:TGK}
\kappa_{xx} = \int_0^{\infty} \langle v_x(0)v_x(t)\rangle dt,
\EE
where $v_x$ is the velocity of the particle in the $x$-direction and the angle brackets denote an ensemble average over particle trajectories.
The original nonlinear guiding center (NLGC) theory \citep{MatthaeusEA03} replaced the velocity in this expression with the guiding center velocity
\BE \label{eq:GCV}
\tilde{v}_x = a v_z(t) b_x[\vec{x}(t),t]/B\zero,
\EE
where $a$ is a constant that connects the guiding center $\tilde{v}_z$ with $v_z$. Using this in \Eq{eq:TGK} yields
\BE
\kappa_{xx} = \frac{a^2}{B\zero^2}\int_0^{\infty} \langle v_z(t) v_z(0) b_x[\vec{x}(t),t] b_x[\vec{x}(0),0] \rangle dt
\EE
which involves fourth-order correlations. The NLGC theories then assume $\langle v_z(t) v_z(0) b_x[\vec{x}(t),t] b_x[\vec{x}(0),0] \rangle \approx \langle v_z(t) v_z(0) \rangle \langle b_x[\vec{x}(t),t] b_x[\vec{x}(0),0] \rangle$,
i.e., that the decorrelation of the particle velocity component $v_z$ (related to parallel scattering) is statistically independent of the decorrelation of the magnetic fluctuation $b_x$.  
The UNLT theory \citep{Shalchi2010} models the fourth-order correlation using a Fokker-Planck approach. 

The resulting theories can be written using a Fourier representation in the form
\BE
\kappa_{\perp} = \frac{a^2 v^2}{3 B\zero^2} \int S_{xx}(\vec{k})T(\vec{k})d\vec{k},
\EE
where $S_{xx} = \langle b_x^{\nostar}(\vec{k})b_x^{*}(\vec{k}) \rangle$ is the power spectrum of $b_x$, $T$ is an effective parallel scattering time, and we have replaced $\kappa_{xx}$ with $\kappa_\perp$ for convenience, which is valid given the assumed axisymmetry. Taking $\kappa_\parallel = \kappa_{zz}=(1/3)v\lambda_\parallel$ the original NLGC theory has
\BE \label{eq:NLGC}
T(\vec{k}) = \frac{1}{\kappa_{\parallel}k_z^2 + \kappa^{\notwo}_{\perp}k_{\perp}^2 + v^2/(3\kappa_{\parallel})}. \quad \text{(NLGC)}
\EE
The UNLT theory takes
\BE \label{eq:UNLT}
T(\vec{k}) = \frac{1}{F(k_z,k_\perp) + (4/3)\kappa^{\notwo}_{\perp}k_{\perp}^2 + v^2/(3\kappa_{\parallel})}, \quad \text{(UNLT)}
\EE
with $F(k_z,k_\perp)= (v k_z)^2 / (3 \kappa_{\parallel} k_\perp^2)$. The improved NLGC (RBD/BC) theory \citep{RuffoloEA12} has
\BE \label{eq:RBDBC}
T(\vec{k}) =\sqrt{\frac{\pi}{2}}\frac{\erfc{(\beta)}}{\sqrt{\sum_ik_i^2\langle\tilde{v}_i^2\rangle}}, \quad \text{(NLGC RBD/BC)}
\EE
with
\BE
\beta = \frac{v^2/(3 \kappa_\parallel)}{\sqrt{2\sum_ik_i^2\langle\tilde{v}_i^2\rangle}},
\EE
and where $\erfc(\cdot)$ is the complementary error function.
This theory takes guiding centers to initially follow magnetic field lines, with assumed ballistic spreading of field lines over the scale $\lpara$, after which field line diffusion is obtained. A distribution of guiding centers would initially have a variance
\BE
\sigma^2_i=\left<\tilde{v}^2_i\right>t^2,
\EE
where $\left<\tilde{v}^2_i\right>$ is the mean square of the guiding center velocity of a particle in the $i$ direction.
Using \Eq{eq:GCV} one can obtain
\BE
\begin{split}
\left<\tilde{v}_x^2\right>&=\left<\tilde{v}_y^2\right>=\dfrac{a^2v^2}{6}\dfrac{b^2}{B\zero^2},\\
~\text{and}~\left<\tilde{v}_z^2\right>&=\dfrac{v^2}{3}\left(1-\dfrac{a^2b^2}{B\zero^2}\right).
\label{guidingcentervelocity}
\end{split}
\EE
Here we have used $b^2=\langle b_x^2 +b_y^2 \rangle=2\langle b_x^2 \rangle$ by axisymmetry and $\langle v_z^2 \rangle = v^2 /3$, which is expected for an isotropic distribution of particle velocities.
All of the above theories require the parallel mean free path, $\lambda_\parallel$, or equivalently $\kappa_\parallel$, as
an input ($\lambda_\parallel = 3 \kappa_\parallel /v$), which in this work we obtain from test particle simulations.
Note that while \Eqs{eq:NLGC}{eq:UNLT} imply a nonlinear integral equation for $\kappa_\perp$, \Eq{eq:RBDBC} can be evaluated directly. 
In this work we take $a^2 = 1/3$ which has previously been shown to work well, but has only recently found
theoretical explanation \citep{Shalchi2019}
\citep[alternatively, there is a recent theory for NLGC and UNLT that does not require $a^2$;][]{Shalchi21}.
This choice leads to a limitation of the RBD/BC theory in that it
cannot be applied for $b/B\zero > 1/a = \sqrt{3}$, because the expression for $\langle\tilde{v}_z^2\rangle$ in \Eq{guidingcentervelocity} does not make sense. 
(Actually, $b/B_0\gg1$ is generally an unreasonable choice for the case of transverse turbulence as considered by the NLGC and UNLT theories, because as noted earlier, variance anisotropy is expected for a strong large-scale field with $B_0\gtrsim b$.)
In evaluating the above theories, for convenience we can take $S(k_\perp,k_z) = (S_{xx} + S_{yy})/2$ using axisymmetry.

Recently, \cite{Shalchi2019} proposed another expression that includes the FLRW limit for large $\lambda_\parallel$
\BE \label{eq:composite}
\frac{\lambda_\perp}{l_\perp} = \frac{9l_\perp}{16\lambda_\parallel}\left(\sqrt{1 + \frac{8 D \lambda_\parallel}{3\lperp^2}} -1 \right)^2,
\EE
where they used $\lperp$ and $\lpara$ as {\it bend-over scales} (such as $\lambda$ in Equation [\ref{eq:A}]). This expression interpolates between a
\citet{RechesterRosenbluth78}
limit and the FLRW limit as a function of $\lambda_\parallel$. 
However, \citet{Shalchi2019} noted that in principle 
$\lperp$ and $\lpara$ can also be interpreted as the {\it integral (correlation) length scales}, as defined in Section~\Ref{model}. 
In comparison with our test particle simulations, the use of integral scales gives much closer agreement than bend-over scales, so we adopt that usage
in Equation (\ref{eq:composite}) and refer to this as the {\it composite} model.

We will also compare other theoretical and simulation results with the simple field line random walk (FLRW) theory for perpendicular diffusion, which considers that particles follow field lines with no backscattering. We can write this prescription as
\BE \label{eq:FLRW}
\kappa_\perp = \langle|\mu|\rangle vD
= \frac{v}{2} D,
\EE
where $D$ is the perpendicular magnetic field line diffusion coefficient defined in \Sec{sec:FLT}. This result might be expected to apply when $\RL$ is small, or in the limit of large $\lambda_\parallel$.


\section{RESULTS}

As detailed in \Sec{sec:simulations}, we consider the cases $\lpara/\lperp=1,$ 3, and 10. We express the diffusion coefficients
in terms of the parallel ($\lambda_\parallel$) and perpendicular ($\lambda_\perp$) mean free paths. We find for small values of $b/B\zero$ that $\lambda_\parallel$ becomes very large and does not converge (when obtained via \Eq{eq:running}) within
practical simulations. At the same time, $\lambda_\perp$ is highly oscillatory with time. For this reason we have used primarily moderate values $b/B\zero=0.3,$ 0.5, and 1, except where we study the dependence on $b/B\zero$. For reference, our standard parameters correspond to a range of magnetic Kubo numbers $0.3 \le R \le 10$.

\subsection{Parallel Diffusion Coefficients}
\begin{figure}
\begin{center}
	\includegraphics[width=1.0\linewidth]{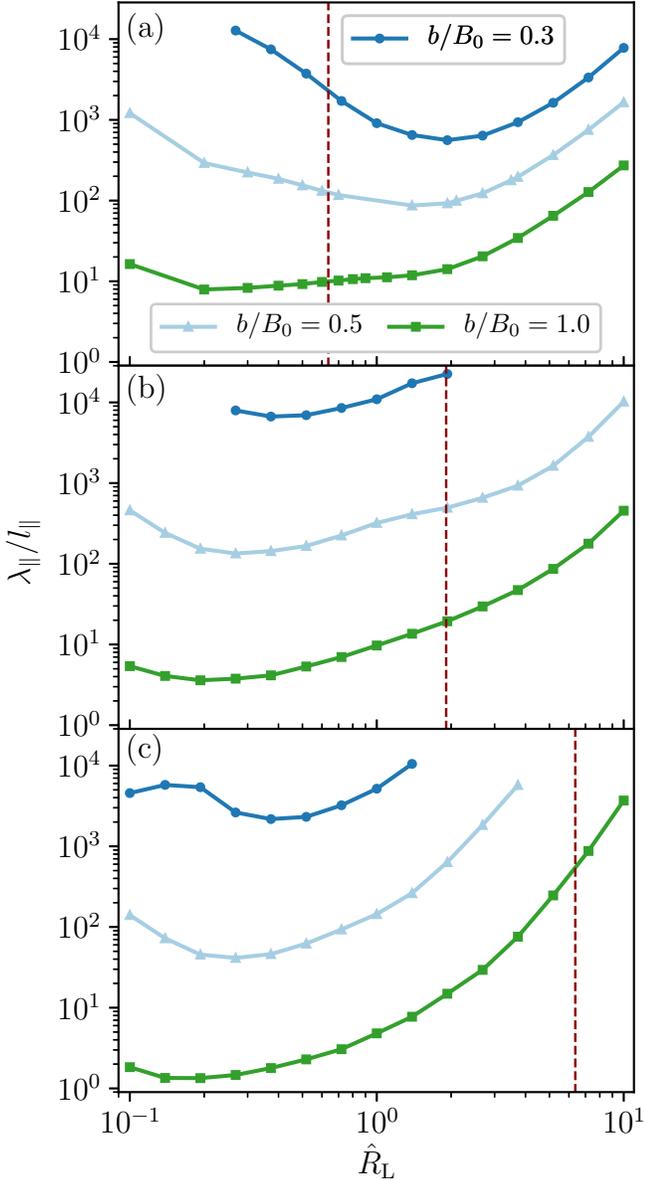}
    \caption{The test particle parallel diffusion coefficient in NRMHD turbulence for (a) $\,\lpara/\lperp=1,$ (b) $\,\lpara/\lperp=3,$ and (c) $\, \lpara/\lperp=10$ as a function of $\HRL=\RLL$ for various $\dbob$. 
    Note that some points, where $\lambda_{\parallel} \gtrsim 3\times 10^4$, are not plotted, as asymptotic parallel diffusion was not reached within the maximum time of the test particle simulations. 
    There is no resonant scattering for $\HRL$ below a cutoff (vertical dashed lines), which is a special feature of the NRMHD model, as is the {\sf U}-shaped dependence of $\lambda_\parallel$ on $\HRL$.
    This work will examine whether theories of perpendicular diffusion remain accurate when resonant scattering is weak or absent.
    }
    \label{fig:kappa_para1}
\end{center}
\end{figure}

In \Fig{fig:kappa_para1} we show parallel diffusion coefficients from test particle simulations for $b/B\zero=0.3,$ 0.5, and 1.0 and $\lpara/\lperp=1$, 3, and 10 as a function of $\HRL=\RLL$, where we have expressed the coefficients in terms of the parallel mean free path, $\lambda_{\parallel} = 3 \kappa_{\parallel}/v$. We also made simulations with $b/B\zero=0.1$, but asymptotic diffusion was not obtained within the total simulation time for the range of parameters plotted here. The resonance condition $\RL k_z |\mu| =1$ implies that there should be no resonant parallel scattering when $\RL/\lpara < 2/\pi$. We have indicated this cutoff with a dashed vertical line in each figure panel.

\begin{figure*}
\begin{center}
	\includegraphics[width=1.0\linewidth]{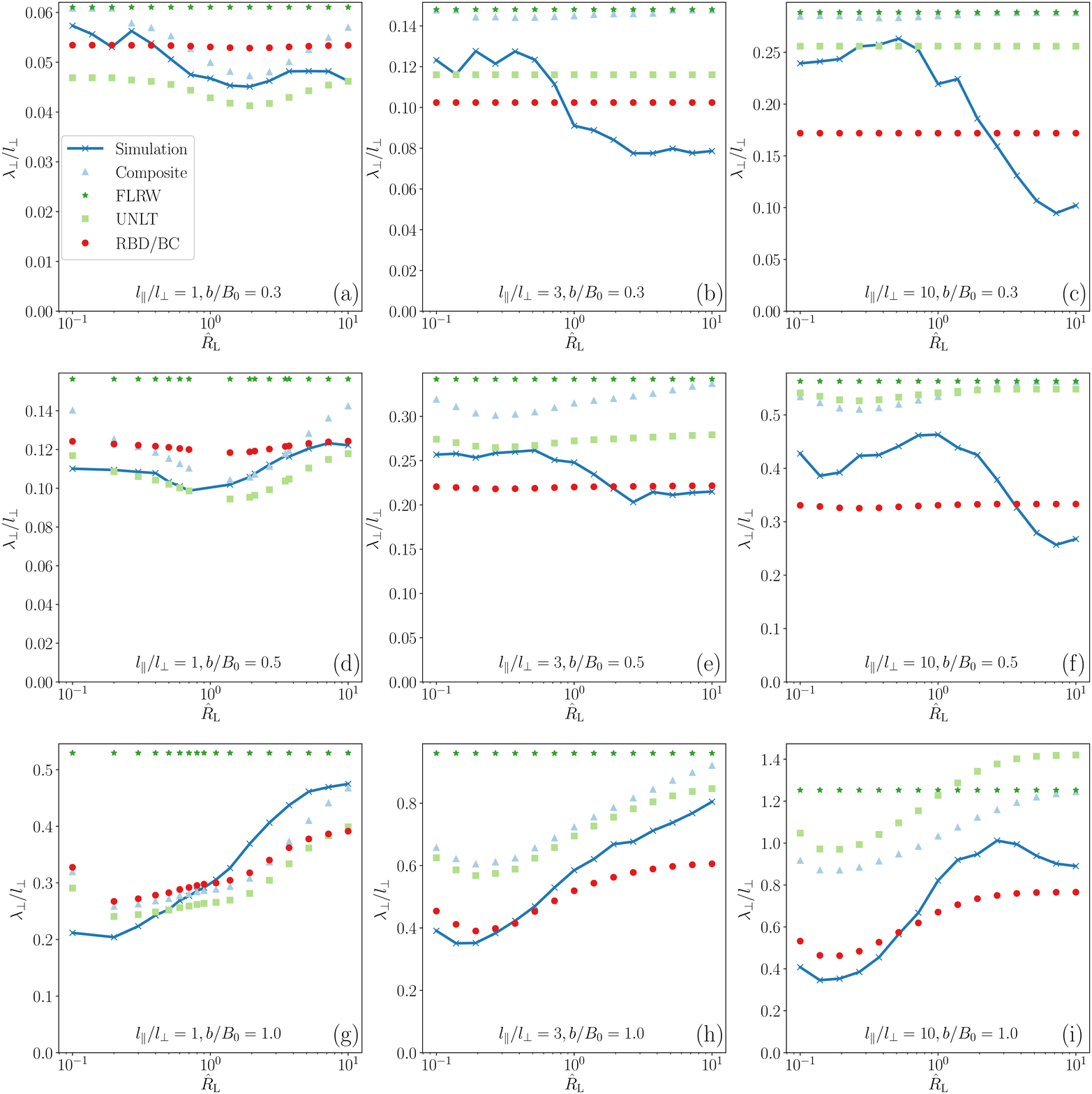}
    \caption{Perpendicular diffusion coefficients from test particle simulations (solid lines) and corresponding theories discussed in the text (points) as a function of $\HRL$ for $\lpara/\lperp=1,$ 3, and 10 (columns from left to right) with $b/\Bzero=0.3,$ 0.5, and 1.0 (rows from top to bottom).
     \label{fig:kaperp}}
\end{center}
\end{figure*}
\begin{figure}
\begin{center}
	\includegraphics[width=1.0\linewidth]{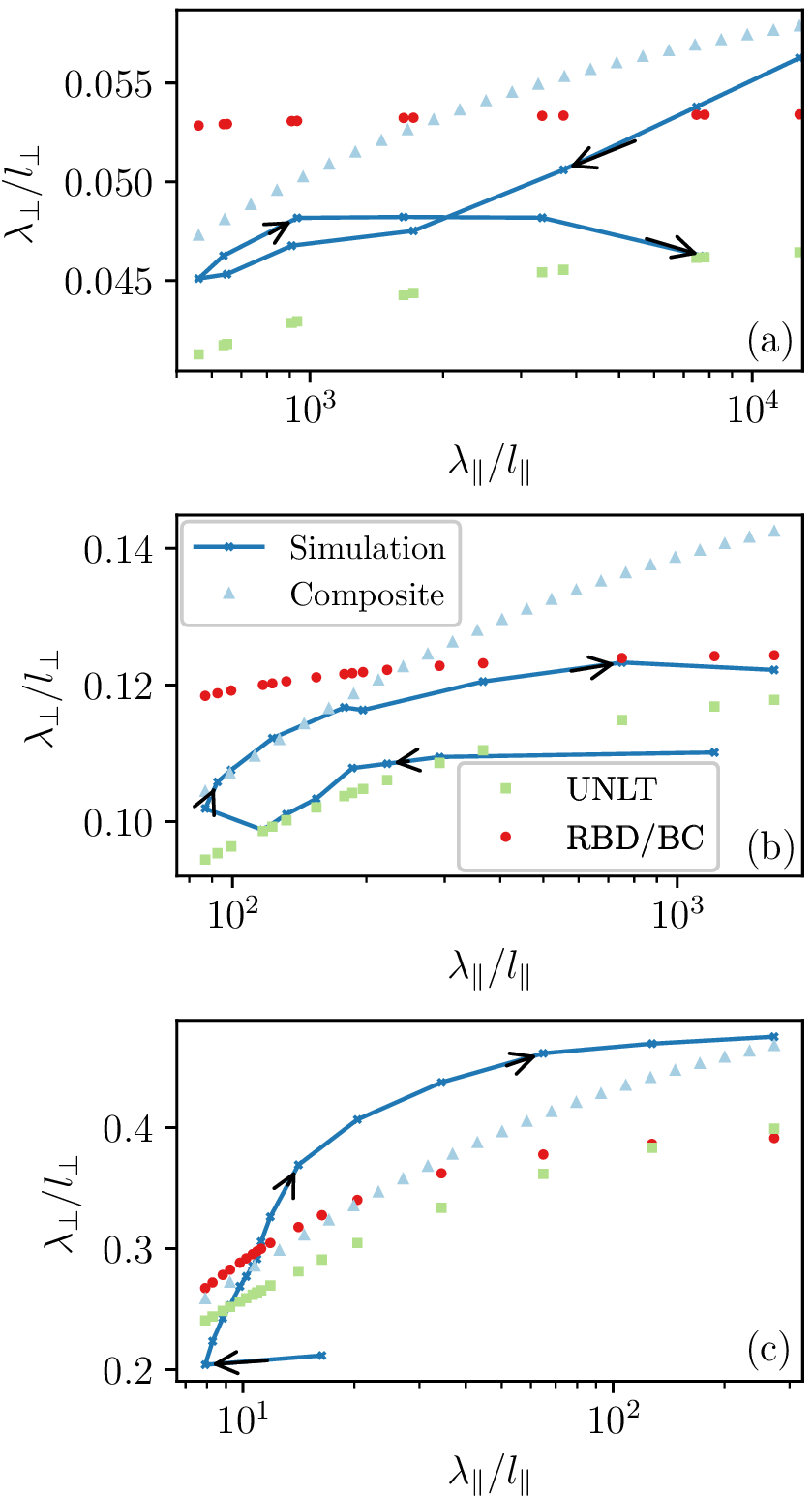}
    \caption{$\lambda_{\parallel}$ against $\lambda_{\perp}$ for simulations (solid) and theory (points) with $\lpara/\lperp=1$. {\bf (a)} $b/B\zero=0.3$; {\bf (b)} $b/B\zero=0.5$; {\bf (c)} $b/B\zero=1$.
    The arrows point in the direction of increasing $\HRL$.
    Despite the hysteresis, which is a special feature of the NRMHD model, 
    for $b/B_0\leq0.5$ the theories 
    match  the simulation results to within 20\%.
    }
    \label{fig:hyste}
\end{center}
\end{figure}

We see in all cases a decrease in $\lambda_{\parallel}$ with decreasing $\HRL$, and then an eventual increase beyond a certain $\HRL$.
For $\lpara/\lperp=1$ we observe the scaling $\lambda_{\parallel}\sim\HRL^2$
for $\HRL \gtrsim 3$, as is predicted by the quasilinear theory of resonant scattering for $\HRL \gg 1$, but the slope is
steeper for the other cases.
In contrast, for $\HRL$ below the cutoff, there is no resonant scattering, only nonlinear scattering, and for
small $\RL$ we find that $\lambda_\parallel$ depends very strongly on $\dbob$.
These parallel diffusion coefficients serve as input for the theories evaluated in the following subsections.

\subsection{Perpendicular Diffusion Coefficients}

In \Fig{fig:kaperp} we compare perpendicular diffusion coefficients from test particle simulations with theoretical
expressions for the same simulations as shown in \Fig{fig:kappa_para1}, using one panel for each simulation.
Where the value of $\lambda_{\parallel}$ is not available from the simulation (i.e., it is not converged) we take $\erfc(\beta)\to1$ in \Eq{eq:RBDBC} and $\kappa_\parallel\to\infty$ in other equations when evaluating the theories.
Note that here (and in several other figures) we use a linear vertical scale, when there is less than about one decade of variation. 
The RBD/BC and UNLT theories and the composite expression match the simulations quite well for $\lpara/\lperp=1$, with a maximum error of about 50\% for the range of parameters considered. For $\lpara/\lperp=3$ and 10, these theories again match the simulation data reasonably well, with the best match depending on $\HRL$ and $b/B\zero$. For $b/B_0\ge0.5$, only the RBD/BC results intersect the simulation results. The perpendicular diffusion does not vary much at a given $\lpara/\lperp$, and all theories agree with the simulations to within a factor of a few. Note that as $b/B\zero$ decreases, the Larmor radius dependence becomes weaker. 
Note also that the theories considered here have no explicit dependence on $\HRL$, and for a single component of turbulence with power in 3D wavenumber space (like NRMHD), they have a dependence on $\lambda_\parallel$ that saturates at large $\lambda_\parallel$. 
Since the parallel diffusion coefficient is very large in the case of $b/B\zero\to0$, the theories give essentially a constant perpendicular diffusion, independent of $\HRL$.
This saturation can explain why $\lambda_\perp$ and $\lambda_\parallel$ may have a very different dependence (or lack thereof) on $\HRL$ \citep[see, e.g.,][]{DundovicEA20}.

\begin{figure}
\begin{center}
	\includegraphics[width=1.0\linewidth]{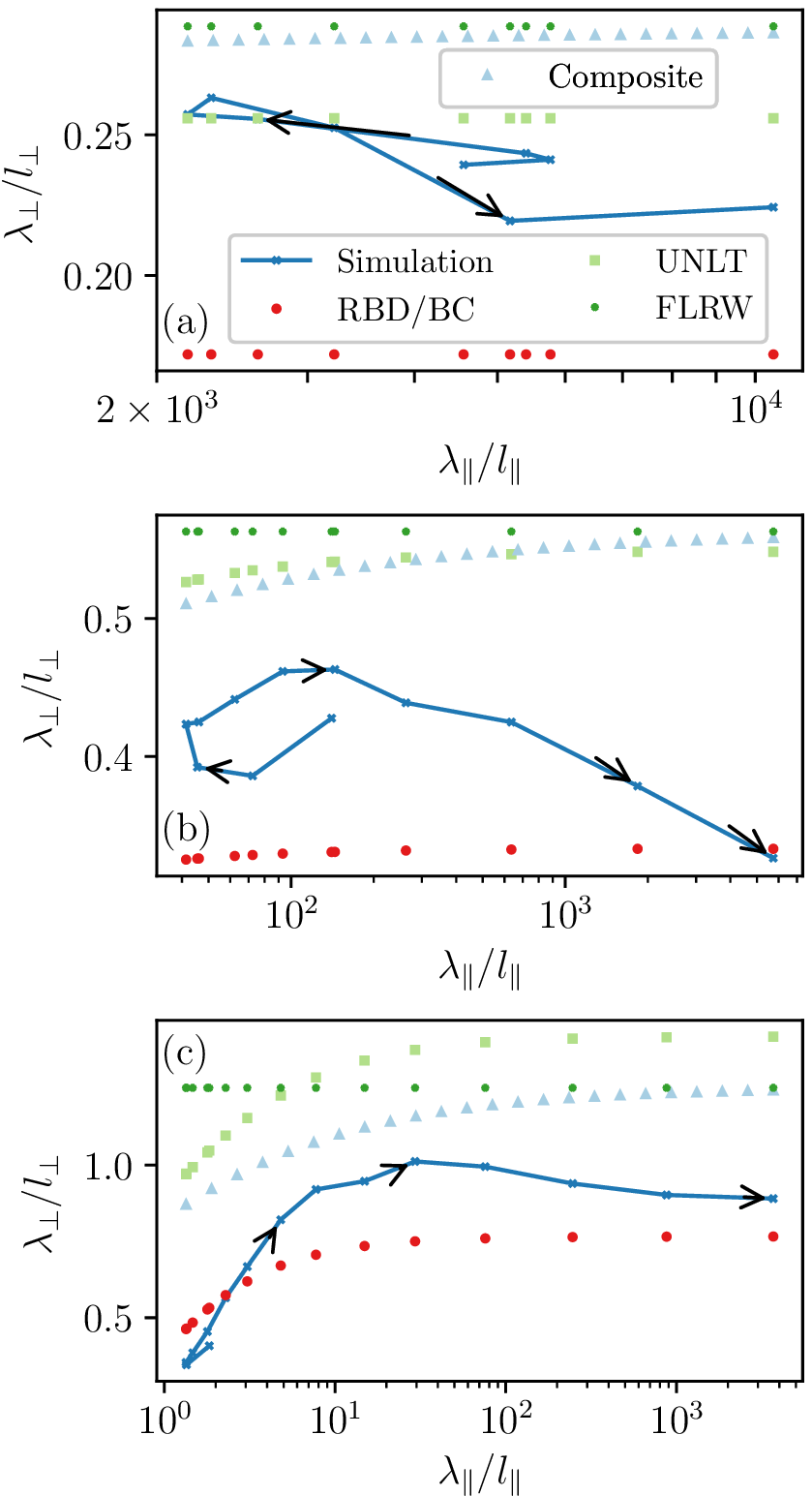}
    \caption{$\lambda_{\parallel}$ against $\lambda_{\perp}$ for simulations (points) and theory (points) with $\lpara/\lperp=10$. {\bf (a)} $b/B\zero=0.3$; {\bf (b)} $b/B\zero=0.5$; {\bf (c)} $b/B\zero=1$.
    In this case the hysteresis effects are weaker for the parameters 
    considered.
    }
     \label{fig:kappa_perp10}
\end{center}
\end{figure}
\begin{figure*}
\begin{center}
	\includegraphics[width=1.0\linewidth]{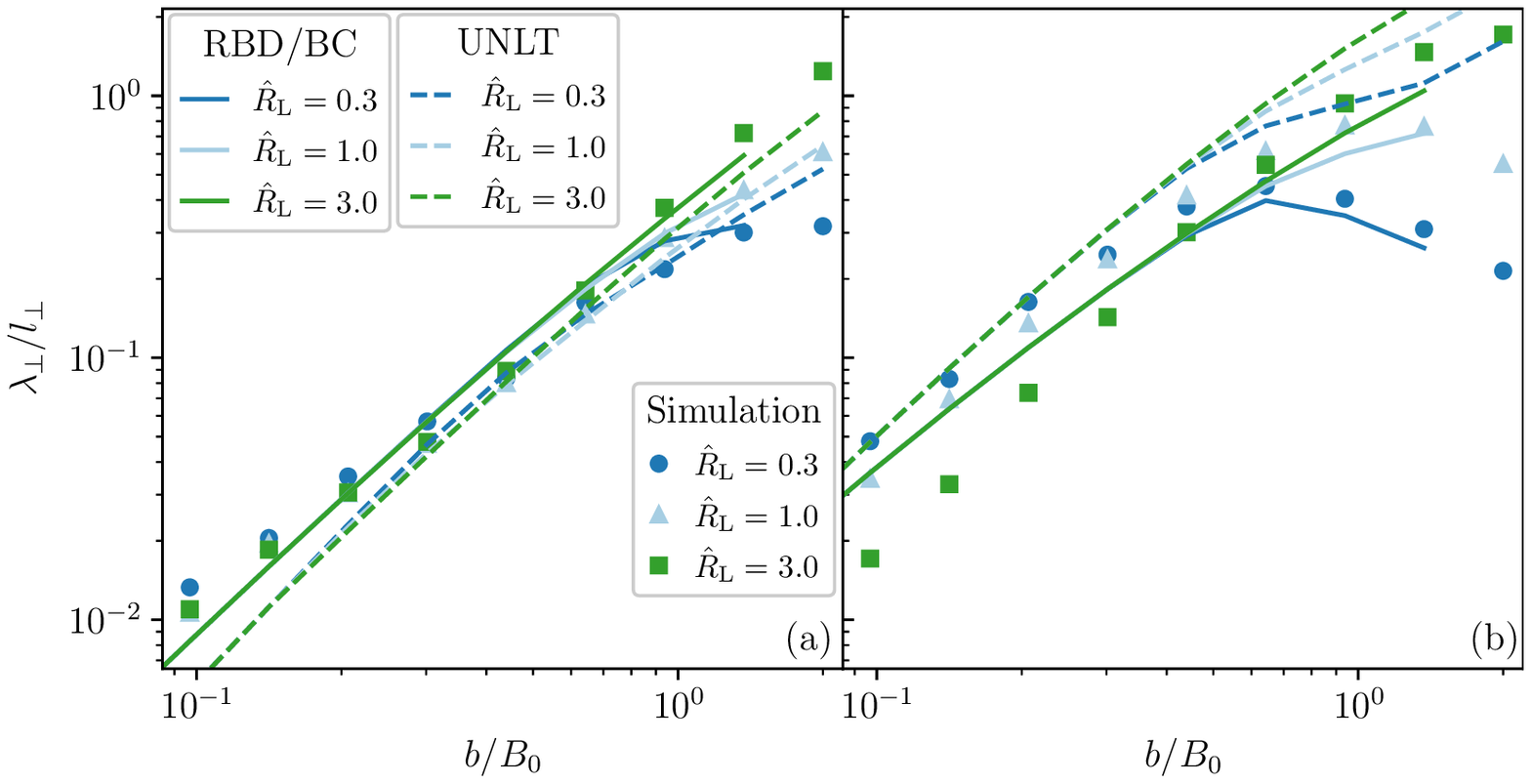}
    \caption{
    The perpendicular test particle diffusion coefficient in NRMHD turbulence (symbols) for (a) $\lpara/\lperp=1$ and (b) $\lpara/\lperp=10$ as a function of $\dbob$ for various $\HRL$, compared with NLGC RBD/BC theory (solid lines) and UNLT theory (dashed lines). 
    The theory results basically remain close to the simulation results, though the theories do not exhibit an $\HRL$ dependence at $\dbob\ll1$ in panel (b).
    }
    \label{fig:kappa_perp_dbob}
\end{center}
\end{figure*}

Due to the reduced parallel scattering at small $\HRL$, the simulations produce a given $\lambda_\parallel$ value for two distinct $\HRL$ (see Figure \ref{fig:kaperp}), which can have different $\lambda_\perp$ values.
This is shown by the hysteresis patterns in \Fig{fig:hyste}, where the solid curves show the variation of $\lambda_\perp$ with $\lambda_\parallel$ for $\lpara/\lperp=1$, with the arrows denoting the direction of increasing Larmor radius.
Actually, the distinct values of $\lambda_\perp$ at a given $\lambda_\parallel$ are usually not very different, which justifies the lack of explicit $\HRL$-dependence as a reasonable approximation. Similar behavior is seen for $\lpara/\lperp=10$, as shown in \Fig{fig:kappa_perp10}. In \Fig{fig:kappa_perp10} we have also shown the FLRW result obtained from the simulated magnetic field lines. 

In \Fig{fig:kappa_perp_dbob} we show the dependence of the perpendicular diffusion on $b/B\zero$ for
$\lpara/\lperp=1$ and 10 and compare it to theory results. 
Note that in panel (a), $b/B\zero$ equals $R$, the magnetic Kubo number, and while in panel (b) $b/B\zero=R/10$.
When $b/B\zero\lesssim 0.5$, $\lambda_{\parallel}$ from the simulations (that
is used as input in the theories) becomes very large, and so the theories depend very weakly on $\HRL$, as seen in the
uppermost panels of \Fig{fig:kaperp}. Then for small $b/B\zero$ the theories depend only on $b/B\zero$ (linearly),
which is different from the simulations for $\lpara/\lperp=10$ (panel (b)). Aside from that difference, the RBD/BC theory works quite well down to
$b/B\zero\approx 0.1$. It also works well for $b/B\zero=1.5$, near the upper limit of the theory $b/B\zero=\sqrt{3}$, and appears to have the right trend towards results beyond this limit.
We also made calculations for UNLT, which shows similar trends in parameter dependence, but provides a rigidity dependence at $b/B\zero\gtrsim1$ that is weaker than the dependence found from RBD/BC or the simulation results.
Both theories somewhat agree with the simulations but the agreement is not perfect. In some cases RBD/BC agrees best but in some
other cases UNLT agrees better.

\subsection{Diffusion in the Extended NRMHD Spectrum}
\begin{figure}
  \begin{center}
    \includegraphics[width=1.0\linewidth]{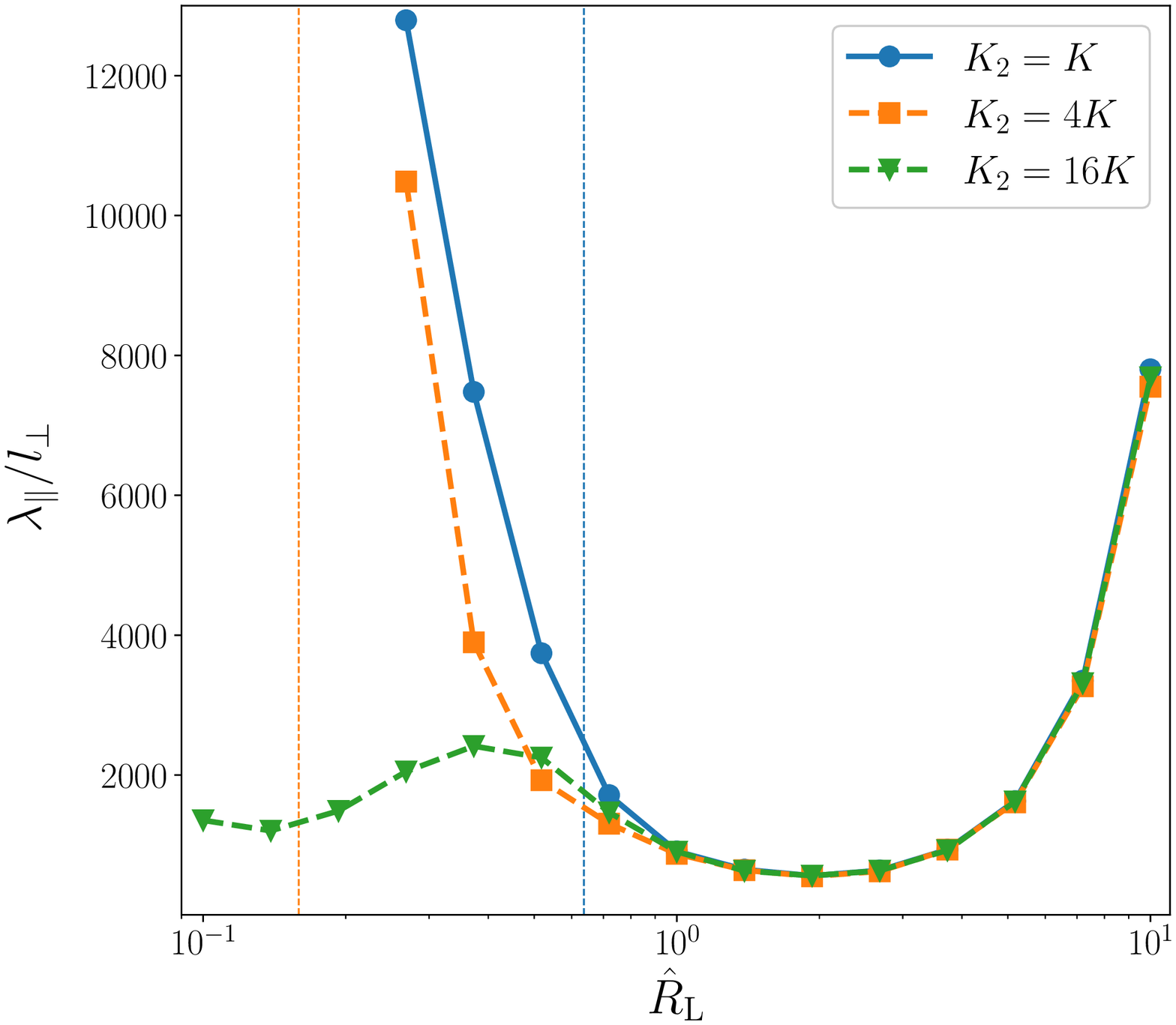}
	\caption{The effect of additional power in the magnetic fluctuation spectrum at high $k_z$, beyond $K$ and up to $K_2$, on the parallel mean free path for $\lpara/\lperp=1$ and $b/B\zero=0.3$. Compared with the standard NRMHD spectrum ($K_2=K$, blue trace), the added power up to $K_2=4K$ (orange trace) or $K_2=16K$ (green trace) reduces the parallel diffusion significantly at small $\HRL$, while it is essentially unchanged at higher $\HRL$.
	Vertical dashed lines indicate the maximum rigidity for which there is no resonant scattering for $K_2=K$ (blue) or $K_2=4K$ (orange). 
    \label{fig:extra_para}}
\end{center}
\end{figure}
\begin{figure}
\begin{center}
    \includegraphics[width=1.0\linewidth]{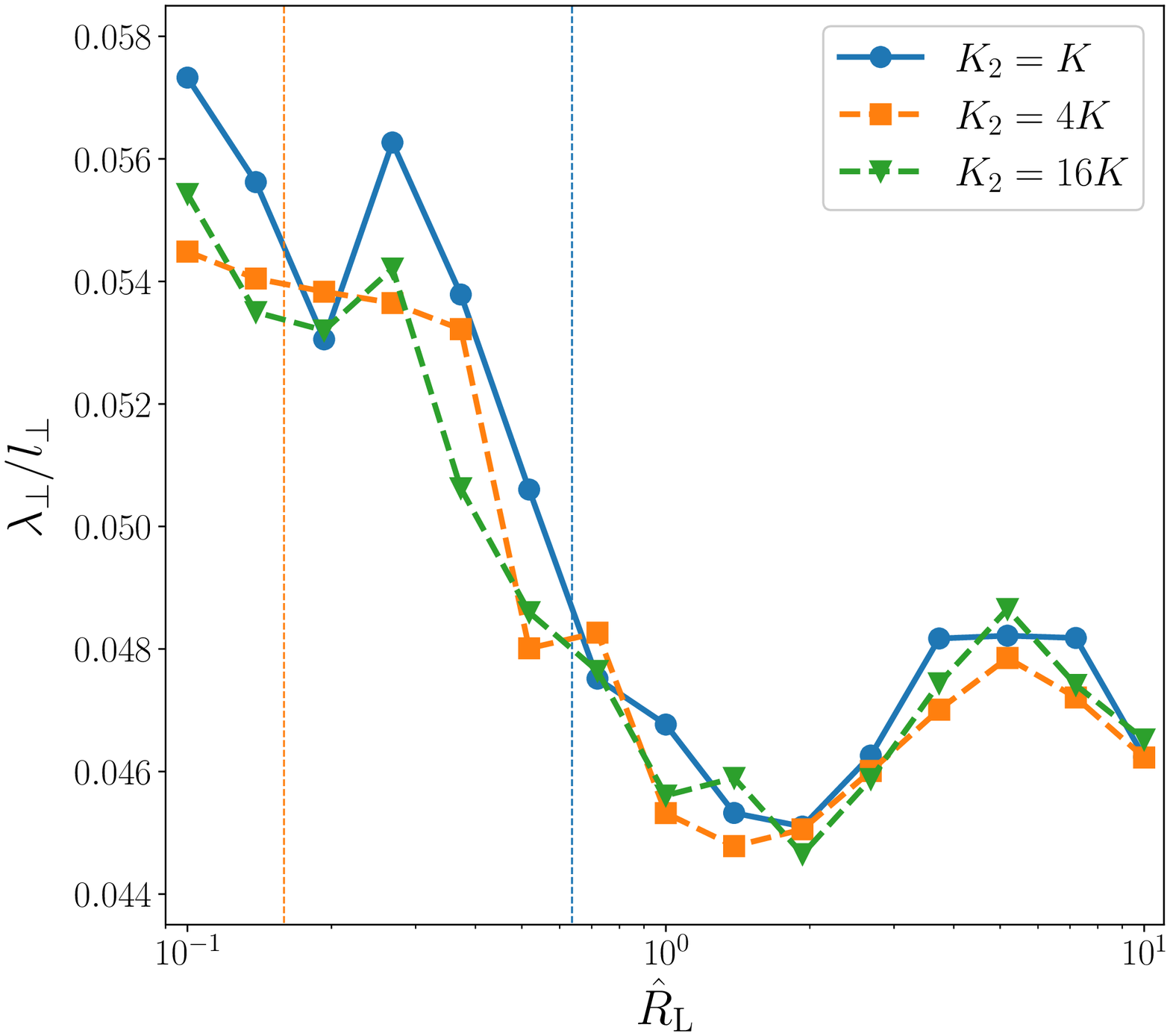}
	\caption{The effect of additional power in the magnetic fluctuation spectrum at high $k_z$, beyond $K$ and up to $K_2$, on the perpendicular mean free path for $\lpara/\lperp=1$ and $b/B\zero=0.3$,  corresponding to the parallel diffusion results in \Fig{fig:extra_para}. Unlike the parallel diffusion, the perpendicular diffusion is hardly affected by the additional power.
    \label{fig:extra_perp}}
 \end{center}
\end{figure}

Here we extend the parallel spectrum to higher $k_z$ according to \Eq{eq:extension}, which leads to finite resonant scattering at lower $\HRL$ than in the previous subsections. In \Fig{fig:extra_para}, where we have taken $\lpara/\lperp=1$ and $b/B\zero=0.3$, 
we show simulation results for the parallel mean free path for the standard NRMHD spectrum ($K_2=K$, blue trace), and when there is additional power up to $K_2=4K$ (orange trace) or $K_2=16K$ (green trace). 
The maximum rigidity for which there is no resonant scattering is now $R_L=1/K_2$, and is indicated in the Figure for $K_2=K$ (blue dashed line) and $K_2=4K$ (orange dashed line); for $K_2=16K$ there is some resonant scattering for the entire rigidity range considered here.
From \Fig{fig:extra_para} we can see that
the additional power has a significant effect on the parallel mean free path for smaller $\HRL$, whereas there is essentially no effect for larger $\HRL$. In contrast, the perpendicular mean free path is hardly affected by this as
can be seen in \Fig{fig:extra_perp}. We have experimented with the extended spectrum for other values of $\lpara/\lperp$ and $b/B\zero$. For $b/B\zero=1$ the effect is negligible.
We find that $\lambda_\perp$ is essentially unaffected in all cases, unless the value of $a_1$ in \Eq{eq:extension} is increased to around $0.1$. However, such a large value of $a_1$ corresponds to a significant change in $\lpara$, so we can no longer attribute the effect directly to enhanced resonant scattering.

Interestingly, the RBD/BC theory for the original spectrum still works for the extended spectrum, despite the reduced parallel mean free path in the particle simulation. This is because the theory does not vary strongly with $\lambda_\parallel$ unless $\lambda_\parallel/\lpara \lesssim 10-100$, as can be seen in \Figs{fig:hyste}{fig:kappa_perp10}. The enhanced resonant scattering applied here does not reduce $\lambda_\parallel$ significantly in this region.

\subsection{The Field Line Random Walk Limit}
\begin{figure}
  \begin{center}
      \includegraphics[width=1.0\linewidth]{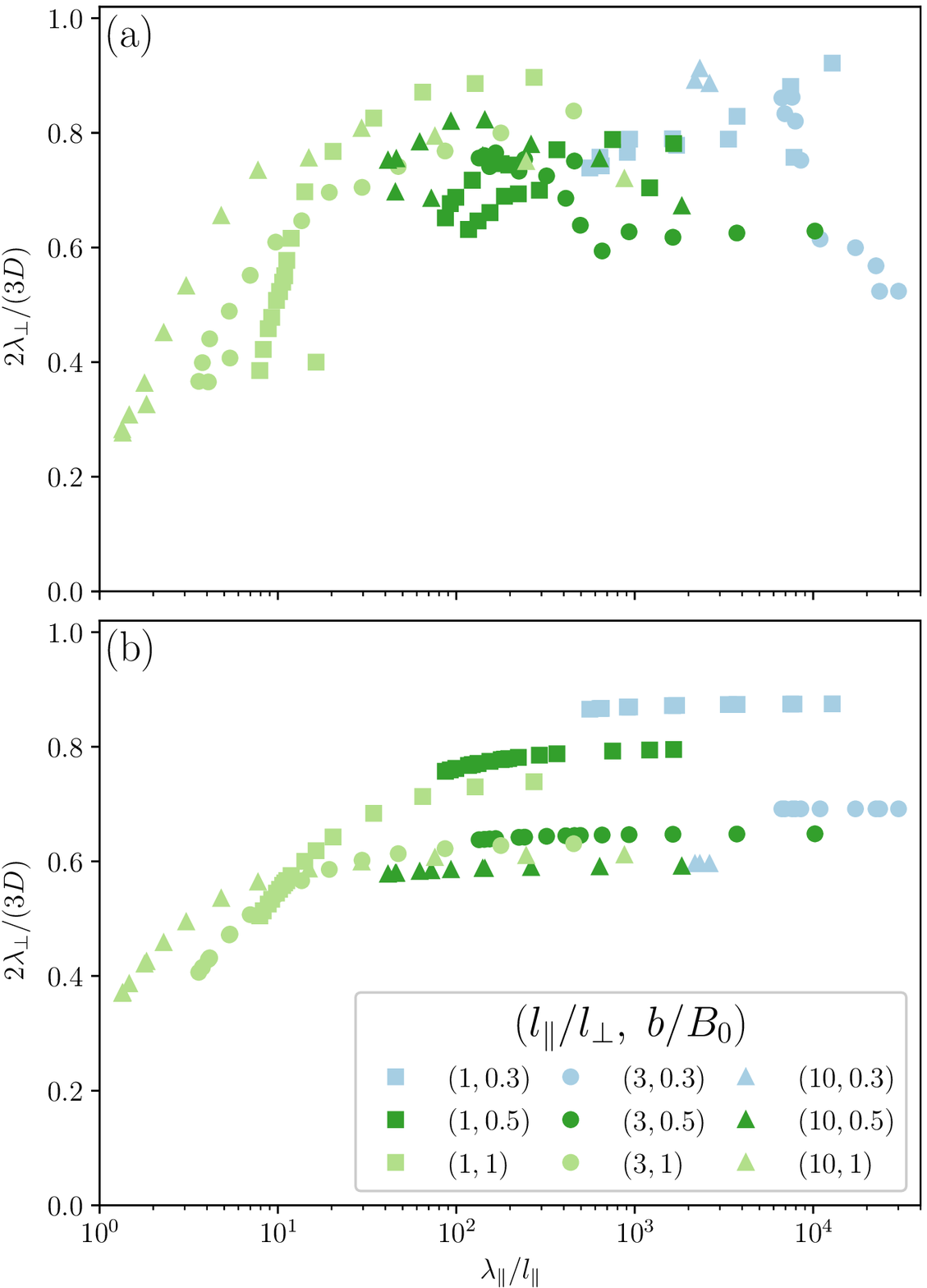}
      \caption{ 
The ratio of $\lambda_\perp$ from (a) test particle simulations and (b) NLGC RBD/BC theory to $3D/2$, where $D$ is the corresponding perpendicular magnetic field line diffusion coefficient obtained from simulation. Each point corresponds to a simulation result and the points are plotted as a function of $\lambda_\parallel$; the symbol color indicates $b/B\zero$ and the shape indicates $\lpara/\lperp$ as specified in the legend.
	The simple FLRW expression yields $1$ for this ratio, overestimating $\lambda_\perp$ from the simulation or NLCG RBD/BC results by at most about a factor of 4.     
      }
    \label{fig:FLRW_SIM_RBDBC}
  \end{center}
\end{figure}

In previous studies, the simulated perpendicular diffusion coefficient has rarely matched the FLRW limit at small $\HRL$ or for large $\lambda_\parallel$, as might be expected. The FLRW limit implies that $\lambda_\perp = (3/2)D$ or $(2/3)(\lambda_\perp/D)=1$. 
We have shown the FLRW limit in \Fig{fig:kaperp} for individual parameter sets, and here in \Fig{fig:FLRW_SIM_RBDBC}(a) we plot $(2/3)(\lambda_\perp/D)$ as a function of $\lambda_\parallel$, combining simulation results for all sets of parameters. 
We see that many of the points are at $0.6-0.8$ with some close to $1$ and a few below $0.5$. The FLRW limit is therefore a reasonable approximation to the perpendicular diffusion coefficient.  FLRW theory overestimates $\lambda_\perp$, but is usually within a factor of 2-4 from the simulation results for NRMHD.
Note, however, that much greater discrepancies were reported from a study for the 2D+slab model of magnetic turbulence \citep{RuffoloEA2008}.
The RBD/BC theory results for NRMHD are also quite close to and always lower than the FLRW limit, as shown in \Fig{fig:FLRW_SIM_RBDBC}(b) for the same parameters as for the simulation points. A similar range of ratios is seen, which is consistent with the closer agreement between RBD/BC results and the simulation results.

\section{Discussion}

The NRMHD model of magnetic turbulence, for $\lpara/\lperp\sim10$ and $b/B_0\ll1$, can be applied to the same physical situations for which RMHD is used, e.g., laboratory plasmas for fusion research and plasma in solar coronal loops.  
More generally, for any parameter values, the model has the special characteristics of a resonance gap above a certain cutoff in Larmor radius (i.e., in rigidity) and a complete lack of resonant scattering below the cutoff, leaving only nonlinear scattering effects. 
Thus the NRMHD model provides a distinct test of the general applicability of theories of particle transport.

Specifically, this model results in a long parallel mean free path with a {\sf U}-shaped profile, rising at both high or low rigidity (Figure \ref{fig:kappa_para1}).  
Indeed, such a profile has previously been inferred from observational data on the parallel transport of energetic charged particles in the solar wind \citep{BieberEA94,GloecklerEA95,Droege00}.
For NRMHD turbulence, parallel transport in the high-rigidity regime is dominated by resonant scattering, while in the low-rigidity regime it is dominated by non-resonant scattering.
Meanwhile, all the theories considered here treat $\lambda_\perp$ as a function of $\lambda_\parallel$ with no explicit dependence on the particle rigidity $\HRL$.
We have examined whether such a framework can accurately describe $\lambda_\perp$, and whether this really depends mainly on $\lambda_\parallel$ regardless of the physical mechanism for parallel scattering.

Here we have performed test particle simulations of energetic charged particles in the noisy RMHD turbulence model and compared perpendicular diffusion coefficients with those provided by a number of theories for a wide range of parameter values. 
In particular it has been demonstrated that the NLGC (RBD/BC) theory, previously applied to two-component 2D+slab turbulence \citep{RuffoloEA12}, is also quite applicable to the NRMHD turbulence model. The RBD/BC theory is in agreement with test particle diffusion coefficients to within about a factor of 2 over the range of parameters considered and provides the best overall agreement to simulations among the theories considered. It also has the advantage over the original NLGC and UNLT theories in that it only involves integration, rather than solving an integral equation. 
For NRMHD, with a paucity or lack of resonant scattering, the parallel mean free path was found to be typically quite large, yet the perpendicular diffusion coefficient exhibits only a weak dependence on both this and the particle Larmor radius, varying only by about a factor of two over two orders of magnitude in $\HRL$. Then a simplification of the RBD/BC theory, taking $\erfc(\beta)\to 1$ in \Eq{eq:RBDBC}, would provide a good approximation in many cases. 
Indeed, the key feature of all these theories that allows them to provide a reasonable description of the simulation results is that $\lambda_\perp$ tends to a constant value at high $\lambda_\parallel$.

For both the RBD/BC and UNLT theories we have adopted $a^2 = 1/3$, as in the original NLGC theory \citep{MatthaeusEA03}. 
There have been several comments in the literature about whether and when $a^2$ should have a different value
\citep[e.g.,][]{ShalchiEA04-WNLT,Shalchi2010}.
One can see in \Fig{fig:kaperp} that 
the RBD/BC are distributed roughly symmetrically about the simulation results, and so a 
value of $a^2$ different from 1/3 would worsen the agreement with our simulation results.
Moreover, a theoretical explanation for $a^2=1/3$ has recently been presented \citep{Shalchi2019}. 
These considerations justify the general use of $a^2=1/3$.

All the theories considered match the results of test particle simulations in NRMHD turbulence to within a factor of a few (more explicitly, the ratio of theory to corresponding simulation value is in almost all cases between $1/3$ and $3$). The simplest theory to apply is the FLRW limit, for which we have shown the accuracy in detail in \Fig{fig:FLRW_SIM_RBDBC}(a). 
(Here we obtained the field line diffusion coefficient $D$ via test field line simulations, but the analytical expressions given by \cite{RuffoloMatthaeus13} have been shown to be very accurate in practice \citep{SnodinEA13b}, provided the magnetic Kubo number $R=(b/B_0)(\lpara/\lperp)$ is not too large.) 
That the FLRW limit works so well is related to the very long parallel mean free paths in NRMHD turbulence.
In previous work for turbulence that contains a slab component, the test particle perpendicular diffusion has usually become much smaller than the FLRW limit as the Larmor radius is reduced \citep[e.g.,][]{RuffoloEA2008}. 
In the present work, for $b/\Bzero=1$ we see a hint of this kind of behavior, but not in other cases. 
Even the related work of \cite{ShalchiHussein2014}, for NRMHD turbulence, does not appear to show such close agreement with the FLRW limit (though it also shows a weak variation of the perpendicular diffusion coefficient with Larmor radius). Interestingly, we note that a recent theoretical result \citep{Shalchi21} for two-dimensional turbulence in the limit of large $\lambda_\parallel$ yields a reduced FLRW expression, so that $2\lambda_\perp/(3D) \approx 0.7136$, which is similar to values shown in \Fig{fig:FLRW_SIM_RBDBC}.

The simple composite expression performs better than the FLRW limit, and about as well as the UNLT theory.
If $\lambda_\parallel$ can be obtained, this provides a convenient estimate of $\lambda_\perp$. We note
that in order to achieve this accuracy it is important to interpret the length scales in the composite expression as integral length scales rather than bend-over scales.

The NRMHD model of magnetic turbulence is interesting in that the sharp cutoff at $|k_z|=K$ in the parallel power spectrum corresponds to a significant gap in the range of pitch angles at which resonant scattering can occur.
Furthermore, for $K \RL < 1$ there should be no resonant scattering according to the quasilinear scattering theory \citep[e.g.,][]{Jokipii66}, so we refer to this region as the non-resonant scattering regime (see Figure \ref{fig:orient}). This leads to an unusual feature in our results that the parallel mean free path is non-monotonic in $\RL$.
This might be expected to challenge the theories of perpendicular diffusion, as they lack an explicit dependence on $\RL$, but in fact the simulation results show only weak hysteresis in $\lambda_\perp$ vs.\ $\lambda_\parallel$ and all theories of perpendicular transport
are fairly accurate. 

Interestingly, the minimum parallel mean free path is sometimes found above the non-resonant scattering regime, for example when $b/\Bzero=0.3$ and $\lpara/\lperp=1$, and in most other cases the parallel mean free path seems to decrease with decreasing $\HRL$ well into this regime (Figure \ref{fig:kappa_para1}). 
This issue can also be examined in terms of a fully-resonant scattering regime, as proposed by \cite{ReichherzerEA20}. This regime is a subset of the regime with some resonant scattering ($K\RL>1$) for which particles are expected to be able to scatter across the resonance gap, $|\mu|<1/(KR_L)$, to access the full range of pitch angles. Their definition of the resonant scattering regime corresponds to
\begin{equation}
\RL \frac{\Bzero}{\sqrt{b^2 + \Bzero^2}} \gtrsim \frac{2}{K}\frac{\Bzero}{b},
\label{eq:Reich}
\end{equation}
where the factor $\Bzero /\sqrt{b^2 + \Bzero^2}$ has been added here to account for the difference in definition of $\RL$, which they took to be defined with respect to the {\it total} magnetic field strength. For our results this regime consistently starts for $\HRL$ above the value that minimizes $\lambda_\parallel$. For example, in the case of $\lpara/\lperp=1$ and $b/\Bzero=0.5$, Equation (\ref{eq:Reich}) yields a fully-resonant scattering regime $\HRL \gtrsim 2.85$, whereas the minimum in $\lambda_\parallel$ is at about $\HRL=2$.
While this seems to provides a bound for the minimum in $\lambda_\parallel$, it does not exactly specify where the minimum should occur; this is likely to depend on nonlinear resonance effects that we have not considered here.
As in \cite{ReichherzerEA20}, this definition of the resonant scattering regime might also explain why the scaling $\lambda_\parallel \sim \HRL^2$ is only seen for $\lpara/\lperp=1$, but further simulations would be required to confirm this.

A development in the present work was to extend the NRMHD power spectrum by spreading a small fraction of the total magnetic energy to a higher maximum parallel wavenumber, $K_2>K$. This was an attempt to reduce the very large parallel mean free path $\lambda_\parallel$ by introducing finite resonant scattering to those particles where $\RL K < 1$, while at the same time keeping $\lpara$ (and $K$) approximately constant. 
This might be expected to further challenge the theories of perpendicular transport, because they explicitly depend on $\lambda_\parallel$.
This small amount of additional turbulent power indeed lead to a significant reduction in the parallel mean free path in some cases, but the perpendicular diffusion was hardly affected in any case. The RBD/BC theory itself provides a hint as to what is happening. In \Figs{fig:hyste}{fig:kappa_perp10} the theory only varies significantly when $\lambda_\parallel/\lpara \lesssim 10^2$.
In those cases where the extended spectrum reduced $\lambda_\parallel$, it rarely went as low as $\lambda_\parallel/\lpara = 10^2$. Moreover, in the cases where the extended spectrum hardly reduced $\lambda_\parallel$, the original value was already around
$\lambda_\parallel/\lpara \lesssim 10^2$. Then without significantly changing the form of the spectrum, this behavior seems to be quite robust.
This suggests that in more realistic fluctuations, such as those in a RMHD simulation, perpendicular diffusion might also be dominated by the large scale magnetic field structure, and in particular the random walk of magnetic field lines, rather than details of the small scale structure.

\acknowledgements
{This work was partially supported by the Thailand Research Fund (grant RTA5980003) and Thailand Science Research and Innovation (grant RTA6280002).
W.H.M.\ was supported in part by NASA through
the Parker Solar Probe project under subcontract SUB0000165 from Princeton University,
and through HSR grant 80NSSC18K1648.
}


\bibliographystyle{aasjournal}
\bibliography{refs}


\end{document}